\documentclass[12pt,a4paper]{article}
\usepackage[latin1]{inputenc}
\usepackage [dvips]{graphicx} 
\usepackage{cite}      
\newcommand{\be}{\begin{equation}}
\newcommand{\ee}{\end{equation}}
\begin{document}

\title{Analyzing fragmentation of simple fluids with percolation theory}
\author{X. Campi\footnote{campi@ipno.in2p3.fr}, H. Krivine, N.
Sator\\{\small\it LPTMS\footnote{UMR 8626, CNRS Universit\'e de Paris XI},
F-91405 Orsay Cedex, France}\\ and E. Plagnol \\{\small\it IPN, F-91406 Orsay
Cedex, France }} \maketitle

\begin{abstract}
We show that the size distributions of fragments created by high energy
nuclear collisions are remarkably well reproduced within the framework of a
parameter free percolation model. We discuss two possible scenarios to explain
this agreement and suggest that percolation could be an universal mechanism
to explain the fragmentation of simple fluids.
\end{abstract}

\section{Introduction}

Fragmentation phenomena concern a wide diversity of objects in nature,
at many scales of distance and time. A natural question is to
ask what is generic and what is specific in these phenomena. Most
theoretical efforts made to understand fragmentation apply only to
specific objects or, on the contrary, concern simple models with few
links with reality \cite{THEO}.  Moreover, experimental
data\cite{TUR93,BEY95,ISH92,Assiettes} are rather sparse and often the
experimental conditions are ill defined. As a consequence, the
question of the possible existence of universal fragmentation
mechanisms remains an open problem \cite{KUN99}.

 The arguments in favour of the existence of universal mechanisms are
of various orders. For instance, in aggregation phenomena, which can
be considered as the opposite of fragmentation, it is possible to
define universal classes\cite{HER86,JUL87} (Diffusion Limited
Aggregation, Clustering of clusters...) in terms of the initial
conditions (number of seeds) and the motion of the aggregating
particles (ballistic, Brownian). The fractal structure (dimension) of
the aggregation cluster is the fingerprint of these universal
classes. In many fragmentation processes the experimental observation,
over many orders of magnitude, of power law (scale invariant) fragment
size distributions is another possible indication of universal
classes.  In this case, the value of the power law exponent would be
the corresponding fingerprint \cite{TUR93,KUN99}.

Collision induced fragmentation of small fluid drops (atomic nuclei
\cite{FRAG_noyaux}, atomic aggregates \cite{CAMB96,FAR98}, liquid droplets
\cite{MEN??}) is a field of experimental research particularly active because
it offers the best possibilities of complete identification of the
fragmentation products. We present in this paper an analysis of the
fragmentation of atomic nuclei in high energy collisions. We show that random
percolation theory accounts for experimental data without any adjustable
parameter and we discuss two possible explanations for this agreement,
depending if one assumes or not that thermal equilibrium is reached before
fragmentation occurs. It is then suggested that this {\it percolation
type} fragmentation mechanism could be universal for simple fluids, {\it i.e.}
fluids made of structureless particles interacting with short range
potentials.

The structure of the paper is the following. In section 2, a brief description
of the experiment precedes the analysis of data. Section 3 is devoted to the
interpretation of the results. Some final remarks and conclusions are
in section 4.

\section{Analysis of fragmentation data}

\subsection{The experiment}

\begin{figure}
\includegraphics[angle=-0,scale=1]{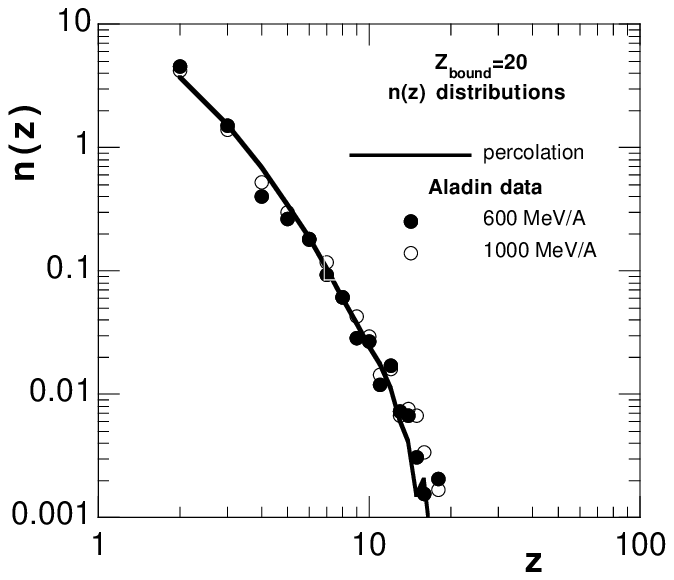}
\includegraphics[angle=-0,scale=1]{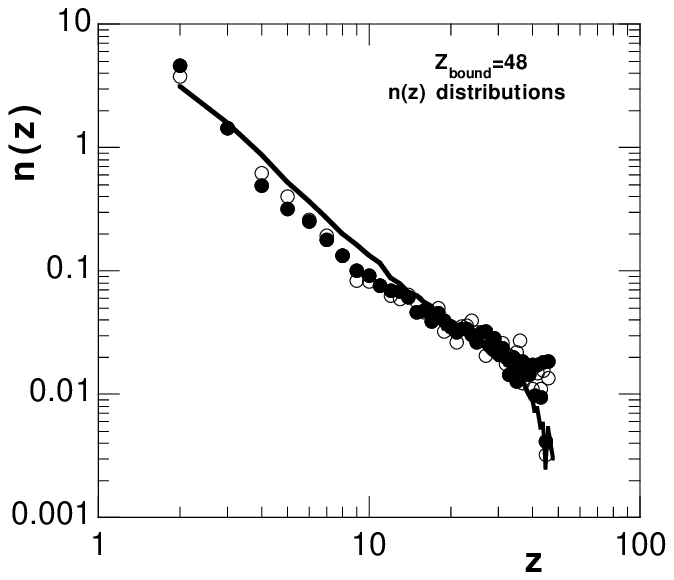}
\includegraphics[angle=-0,scale=1]{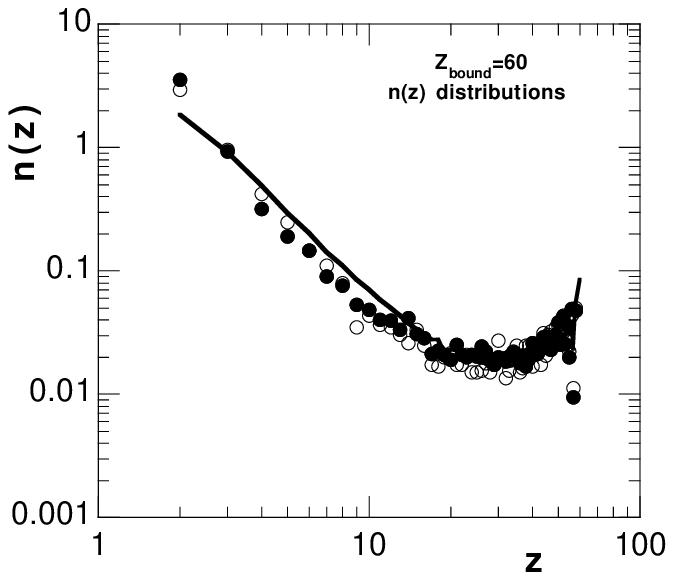}
\includegraphics[angle=-0,scale=1]{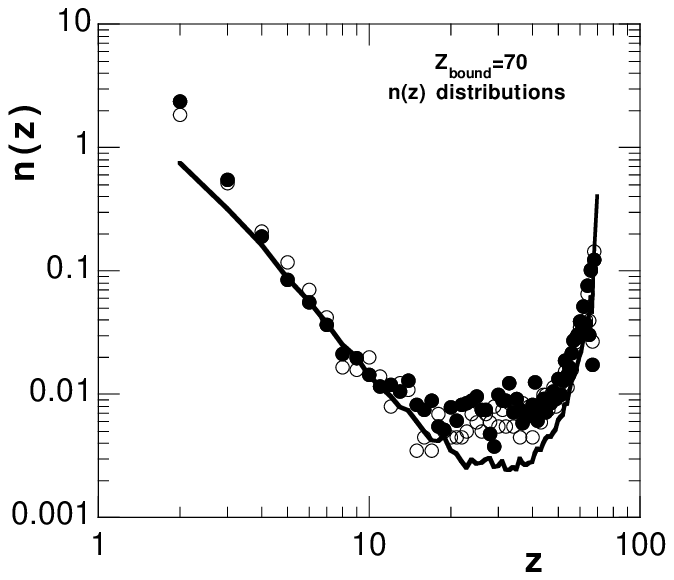}
\caption{\it \label{f:x} Experimental fragment size distributions $n(z)$ at
four values of the control parameter $Z_{bound}$ (see text). The lines
correspond to the percolation calculation. A log-log representation has been
chosen to emphasize the power law behaviour at $Z_{bound}=48$.}
\end {figure}

\begin{figure}[h]
\begin{center}
\includegraphics[angle=-0,scale=1]{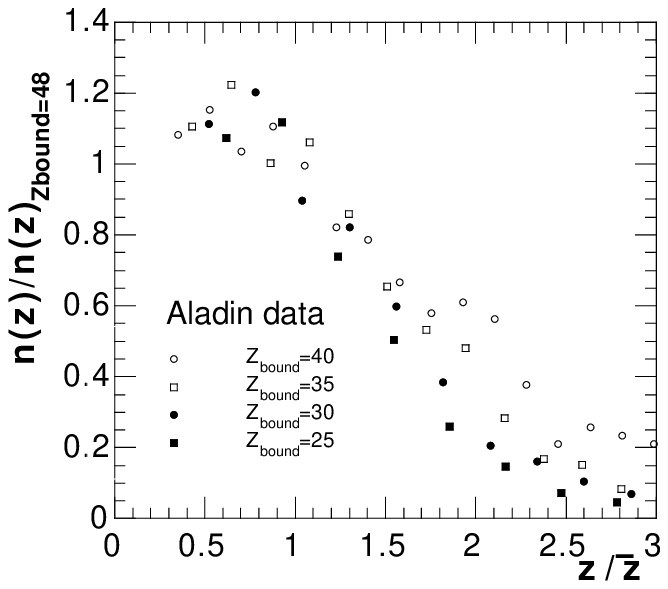}
\caption{\it \label{f:univ} Test of the scaling assumption (Eq. 2), that
 requires that all the ratios $n(z)/n_{c}(z)$  follow the same curve.}
\end{center}
\end{figure}

The experiment was performed by the {\it Aladin} collaboration at {\it
 Gesellschaft für Schwerionenschung} (G.S.I.), Darmstadt \cite{GSI??}. Beams
 of gold ions (Au, $Z=79$) at 600 and 1000 MeV/nucleon incident energies were
 used to bombard targets made of thin copper foils. Data for more than
 $3.10^5$  events were collected. At these high energies\footnote{High energies
 compared with the 8 MeV/nucleon binding energies or the 35 MeV/nucleon Fermi
 energies of nuclei.}, the commonly admitted scenario of the collision is the
 following: The part of the projectile that does not geometrically overlap
 with the target at the point of closest approach is thought to decouple from
 the rest and form a sub-system called the {\it projectile spectator} (PS).
 The size of this system and its excitation energy $E^*$ are therefore
 dependent on the impact parameter. The overlapping part of the system is
 completely vaporized and therefore does not contribute to the formation of
 fragments with nuclear charge $(z)$ greater than one.

\begin{figure}[h]
\begin{center}
\includegraphics[angle=0,scale=0.9,height=6cm]{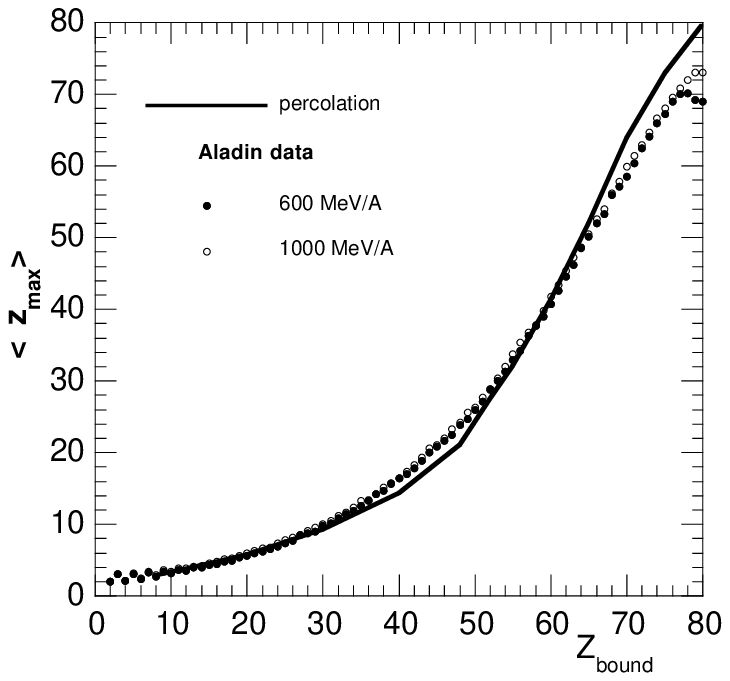}
\includegraphics[angle=-0,scale=.9,height=6cm]{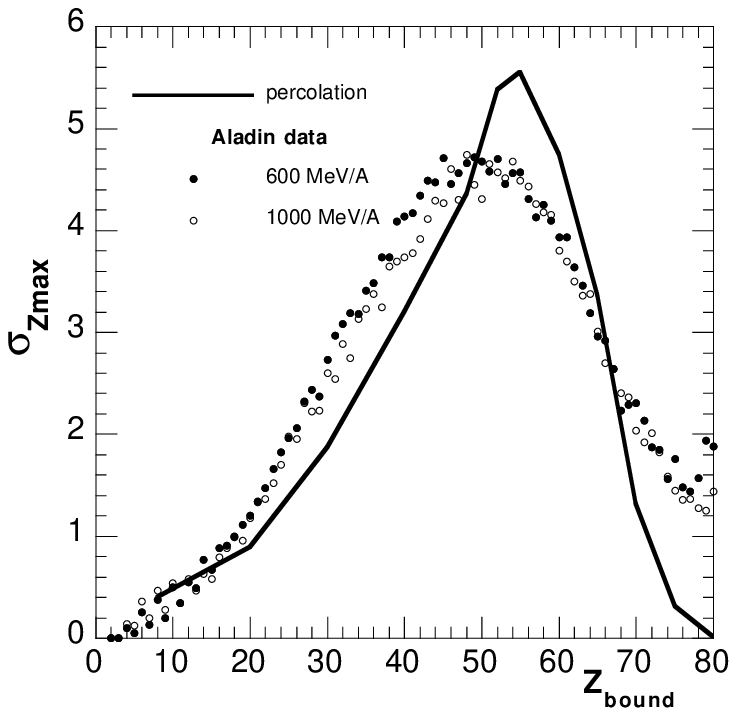}
\caption{\it \label{f:Zmax} Mean value of $z_{max}$ as function of
$Z_{bound}$ for percolation calculation and experimental data from
ref. \cite{GSI??} (left). Fluctuations of $z_{max}$ as function of
$Z_{bound}$ for percolation calculation and experimental data from
ref.\cite{GSI??} (right).}
\end{center}
\end{figure}

\begin{figure}
\includegraphics[angle=-0,scale=1]{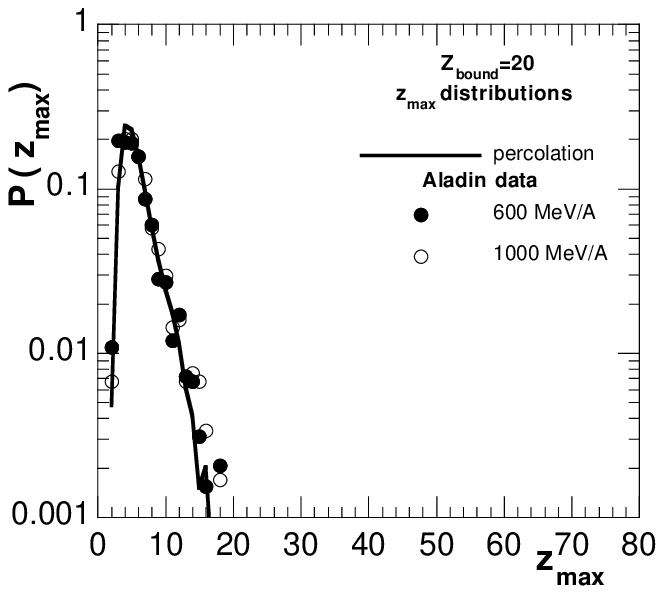}
\includegraphics[angle=-0,scale=1]{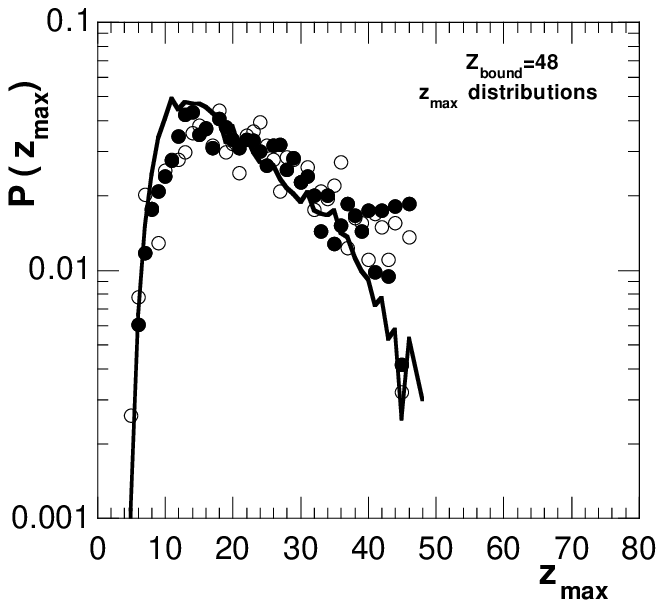}
\includegraphics[angle=-0,scale=1]{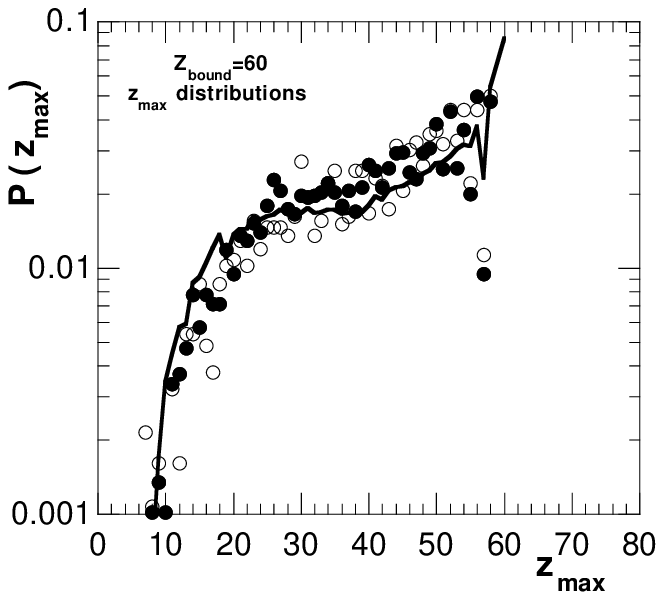}
\includegraphics[angle=-0,scale=1]{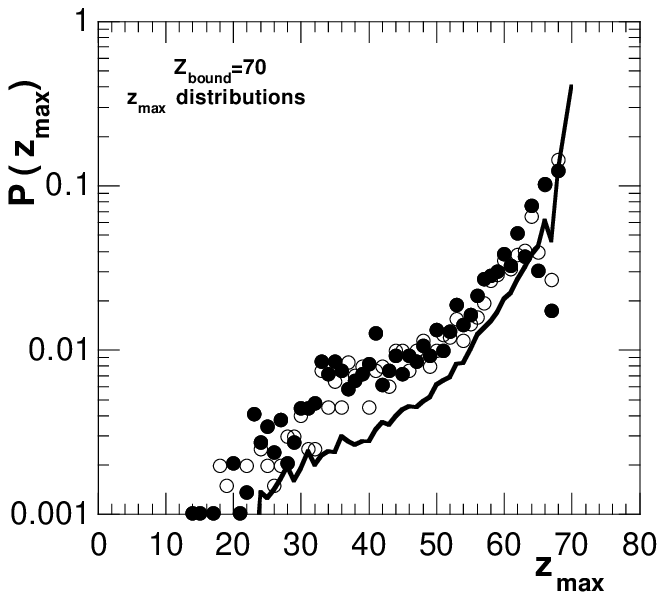}
\caption{\it \label{f:xm} The distribution $P(z_{max})$ of the largest
fragment  at four values of $Z_{bound}$. The circles correspond to
the data from ref. \cite{GSI??} and the lines to the percolation calculation.}
\end{figure}

In the present experiment, the {\it Aladin} device detected with very
high efficiency all spectator fragments (those resulting from the
decay of the unstable PS) with nuclear charge $z>1$. Neither
hydrogen isotopes nor neutrons were detected, event by event, with any
significant efficiency. The initial size of the spectator system ($Z_{PS}$) is
not experimentally measured and can only be inferred from the
comparison with a nuclear reaction model.  Empirically, this size can
be estimated, on average, from the following relation\cite{CAM94}:

\be <Z_{PS}>=25+Z_{bound}-0.004Z_{bound}^2 \label{e1} \ee where $Z_{bound}$ is
equal to the sum of all the charges of products with $z\ge 2$. 
In all the studies of these experiments, the parameter
$Z_{bound}$ is used as control parameter. Notice that $Z_{bound}$, which is
closely related to the impact parameter, decreases when the violence of the
collision increases.

\subsection{Analysis of data}

To analyze the {\it Aladin} data, the following very simple procedure has been
used. For a given value of $Z_{bound}$, the size of the PS was deduced from
eq.\ref{e1} and percolation calculations on a lattice of corresponding size
were performed by varying the bond breaking parameter  (random-bond
percolation). Only those events with the proper $Z_{bound}$ were kept and the
fragment size distribution of these compared to the corresponding experimental
data. In this manner the comparison is parameter free. Notice that by fixing a
value of $Z_{bound}$, the maximum value $z_{max}$ of $z$ is also constrained
by this value.

Figure \ref{f:x} shows, for $Z_{bound}$ values of 20, 48, 60 and 70,
the experimental fragment size distributions. One observes three
different regimes of fragmentation. For large $Z_{bound}$, large
impact parameter and low excitation energy, only a heavy residue and
light particles are produced. This is the ``evaporation'' regime. For
small $Z_{bound}$ (most violent collisions) only small fragments are
produced and no heavy residue is left. The distribution of this
``vaporization'' regime decreases exponentially.  Remarkably, at
$Z_{bound}$ around ${48}$ an intermediate regime exists, for which the
distribution can be fitted by a power law $n(z)\sim z^{-2.2}$. The mean
excitation energies {\it per particle} can be estimated for the
fragmenting systems as a function of $Z_{bound}$\cite{CAM94} : they
correspond to about $<E^*>\simeq 14, 6, 3 ,1$ MeV per particle for
$Z_{bound}=20, 48,60,70$ respectively.

The full lines in figure 1 represent the corresponding distributions obtained
from the percolation calculation. The calculation reproduces quantitatively,
and over several orders of magnitude, the data in the three
regimes\footnote{A close inspection of the data for $Z_{bound}=70$ shows an
excess of fragments produced around $z=30$ due to the presence of fission
events. This is understandably beyond the scope of percolation theory.}. We are
naturally led to associate the experimental power law behaviour at
$Z_{bound}\simeq 48$ to the percolation critical behaviour.

Close to the critical regime, percolation theory predicts typical scaling
properties of the cluster size distributions \cite{STA94}. The size
distributions can be reduced to an universal function, $f(\frac{z}{\bar z})$,
by the following formula :

\be n(z)=n_{c}(z)f(\frac{z}{\bar z}) \label{e2} \ee where $n_c(z)$ is the size
distribution at the critical point, taken here as the size distribution
observed for $Z_{bound}=48$. The ``characteristic size'' $\bar z$ is defined
by :

\be \bar z= m_{3}/m_{2} \ee with $ m_{k}=\sum_{z\ge 2}z^kn'(z)$, where
$n'(z)$ is the mean fragment size distribution obtained excluding,
event by event, the largest fragment. One observes in figure 2 that,
within the scattering of the data, this rule is rather well satisfied,
keeping in mind that this corresponds to the ratio of two quantities
varying over more than three orders of magnitude.

The size of the largest fragment plays in percolation theory the role
of the order parameter. For an infinite system, it is of infinite size
in the percolating phase and finite in the non-percolating one. In a
finite system, this transition is smooth, as illustrated in figure
\ref{f:Zmax}. This figure on the left compares, again as a function of
$Z_{bound}$, the experimental measured size of the heaviest fragment
$z_{max}$ to the one obtained from the percolation calculations. The
figure on the right shows the fluctuations of $z_{max}$,
\be \sigma^2_{zmax}=\frac{<z_{max}^2>-<z_{max}>^2}{<z_{max}>}.\ee
 As expected
\cite{STA80}, these functions show a maximum around the ``critical''
value of $Z_{bound}$. The agreement between experiment and theory is
good, except for the highest values $Z_{bound}$ where the fluctuations
should vanish for $z=79$. Very probably, either contributions of the
target have been included in the data set or miss-identification of
fragments is present. Naturally, the percolation model used here
cannot account for either of these features. In figure \ref{f:xm} one
can see that even the full distribution $P(z_{max})$ is well accounted
for by the calculation.

By any standard, the agreement observed in figures \ref{f:x} to
 \ref{f:xm}, between the experiment and the calculations is very good. In
 the analysis presented here, as already stressed, no adjustable parameters are
 used.

\section{Interpretation of the results}

The choice of percolation theory to analyze the experimental results
is motivated by the following reasoning. A fully microscopic
description of nuclear fragmentation is, a priori, out of scope of
theory. Atomic nuclei behave in their ground state as small drops of
Fermi liquids composed by particles strongly interacting mainly with a
two-body, short range force. This interaction is ill defined at short
distances and the techniques used to solve this complicated many body
problem are not fully under control. The description of collisions,
using transport equations for example, is even more difficult. In view
of the above remarks and with the stated goal of understanding the
universal features of the data, we believe that a more fruitful point
of view is to tackle the problem with the minimum number of assumptions.

 The simplest hypothesis would be to assume the equiprobability of all
partitions of the integer number $Z=79$. This is equivalent to a
``maximum entropy principle'' \cite{AIC84,ENG84}. However, the resulting
${n(z)}$ are always exponentially decaying functions, in contradiction
with experiments.

A step further is to consider topological constraints, by considering
fragments in a 3-dimensional space. Among the infinity of models one can
imagine to make fragments, random-bond percolation seems a good candidate
because of its simplicity and because it retains only the essential
constraints. For example, the shape of the fragments is not assumed a priori.
It turns out that, as shown in section 2, it suffices to reproduce very well
the experimental data. How does one understand this agreement ? One can
imagine at least two scenarios.

In the first scenario, one idealizes the nuclear fluid as an ensemble of
particles connected by bonds. The bond between a pair of particles is active
as long as the magnitude of their potential energy is greater than the
relative kinetic energy. During the collision some of these bonds break
because of the change in the position and/or the velocity of the particles.
The simplest assumption is that bonds are broken randomly, which corresponds
to the usual uncorrelated bond percolation model \cite{STA94}. Particles
connected by unbroken bonds form the fragments. These fragments separate
rapidly, pushed away by the long range Coulomb force between protons.

In a second scenario, one assumes that after the collision phase,
equilibrium thermodynamics applies. The system expands until a
``freezout'' density is reached, at which the fragments cease to
interact by the strong nuclear attractive force and their size
distribution is ``frozen''. The Coulomb force accelerates the
fragments, as before. In order to calculate its distribution, we have
to define first what a ``fragment'' is. In the present context, it
seems natural to call ``fragment'' a self-bound ensemble of particles
\cite{DOR93,PUE99}\footnote{Another possibility is to impose the
stability by particle evaporation only \cite{CAM99}. At the present
level of accuracy, both definitions are operationally very
similar. Definition \cite{CAM99} is equivalent to the ones proposed by
Hill \cite{HIL65} and by Coniglio and Klein \cite{CON80}}. With these
definitions, one finds in the ${\rho - T}$ diagram a percolation line
(sometimes called the Kertész line \cite{KER89}) that separates a
percolating and a non-percolating phases. The line joins the
thermodynamical critical point to the random bond percolation critical
point.  On it the fragment size distribution is a power law $n(z)\sim
z^{-\tau}$, with $\tau\simeq 2.2$. (see fig. 1 and 2 of
ref. \cite{CAM99}). In a small system like an atomic nucleus, rather
than a sharp critical line, one finds a ``critical zone'' on which the
${n(z)}$ approaches this behaviour (see figure 4 of
ref. \cite{CAM97}). The similarity of these calculated ${n(z)}$ with
the experimental ones shown in figure 1 is very obvious. Therefore, by
inspection of the ${n(z)}$ alone, it is not possible to disentangle
between these two scenarios. However,
one could hope that extra information on, for example the fragment
kinetic energies, could indicate if thermalization is present or not.

More generally, the main difficulty in analyzing nuclear fragmentation data is
due to the very small size of the system, which fundamentally limits the
extraction of the universal critical exponents. Indeed, a proper
characterization of the physical process requires, apart from the ${\tau}$
exponent, the determination of the other critical exponents \cite{STA94}
associated with the moments $m_{k}$ of the fragment size distribution. Such
measurements are, in principle, possible for simple fluid systems (i.e. made of
structureless particles subject to short range forces) of larger sizes such as
atomic aggregates or macroscopic pieces of matter, such as liquid drops
\cite{MEN??}.

We consider as very plausible the possibility to observe this {\it percolation
type} fragmentation in simple fluids. Indeed, the arguments that we have
developed to explain the success of percolation theory should apply without
restriction to any system of structureless particles interacting with a short
range potential. In fact, attempts have been made to show experimentally this
behaviour in the fragmentation of hydrogen aggregates \cite{FAR98}. In these
experiments, the multiplicity of fragments $m_{0}$ is used as control
parameter. As a function of $m_{0}$ the mass of the largest fragment and its
fluctuation, evolve qualitatively as expected in percolation theory. However,
in ref.\cite{FAR98}, the data are compared with a percolation system of
improper size and no definite conclusions can be drawn.

 On the theoretical side, we are performing large scale classical molecular
dynamics simulations of a Lennard-Jones fluid, for both the sudden
disassembly of an equilibrated system and for the collisions of drops. We
clearly find in these calculations the fragment size distributions predicted
by percolation theory \cite{CAM20}.

\section{Final remarks}

Other experiments of nuclear fragmentation at high bombarding energies have
been successfully interpreted with percolation theory
\cite{CAM86,KRE93,GIL95,BAU97}. However, at lower bombarding energies (30-50
MeV/nucleon) the agreement is less satisfactory \cite{GANIL,AGO99}. The shape
of the ${n(z)}$ evolves qualitatively as in figure 1, but a closer
examination shows a systematic over-production of intermediate size fragments.
The origin of these discrepancies is still unclear. Different explanations can be considered:

a) the identification of the source of the fragments is more difficult
 at lower collisionnal energies where the separation between the
projectile, the target, or the fused system is not as clear as in the case of
 high energy collisions, leading thus to possible contaminations between these
different sub-systems.

b) different fragmentation mechanisms could result from the smaller relative
velocities between projectile and target, particularly in the case of fusion
at small impact parameter, inducing possible compression effects.

These discrepancies, more generally, could just signal the fact that the
assumption that fragmenting nuclei behave as ``simple fluids'' breaks down at
these lower incident energies.

Nuclear fragmentation experiments are often analyzed with the
so-called Statistical Multifragmentation Models \cite{SMM,SMM1}. In
brief, these models deal with the equilibrium thermodynamics of
ensembles of {\it spherical} drops of {\it nuclear matter} confined in
a ``freezout'' volume. Drops interact with each other only by the long
range Coulomb force and their internal partition function is taken
from empirical mass formulas or from experiment. These
models\footnote{These models have been also implemented to study the
fragmentation of atomic clusters \cite{Gross-Hervieux}.} are successful
in describing the above mentioned low bombarding energy experiments
when fixing the size, the density and the excitation energy of the
fragmenting source. For the high energy {\it Aladin} data similar
analysis have been performed but with a larger set of input
parameters\cite{KRE93}.

 The use of percolation theory concepts is however more comprehensive while
much easier to handle. It provides an excellent agreement with the data
without requiring any adjustable parameter.

We hope that the present results will encourage both
the theoretical and the experimental studies of the fragmentation of simple
fluids. The fragmentation by collisions of very large aggregates seems
particularly promising, because it combines the experimental possibility to
detect  fragments together with a reduction of finite size and surface
corrections effects.

\vskip 0.5cm

We thank the {\it Aladin} collaboration at G.S.I. for allowing us to
use their experimental data.

\end{document}